\documentstyle[aps,12pt]{revtex}
\begin{document}
\renewcommand{\thesection}{\Roman{section}}
\baselineskip 20pt
{\hfill PUTP-96-35}
\vskip 3cm
\begin{center}
{\large\bf Gluon fragmentation to $^3\! D_J$ quarkonia}
\end{center}
\vskip 7mm
\centerline{Cong-Feng Qiao ~and~ Feng Yuan}
\centerline{\small\it Department of Physics, Peking University, 
            Beijing 100871, People's Republic of China}
\centerline{\small \it and China Center of Advanced Science and Technology
(World Laboratory), Beijing 100080, People's Republic of China}
\centerline{ Kuang-Ta Chao}
\centerline{\small\it China Center of Advanced Science and Technology
(World Laboratory), Beijing 100080, People's Republic of China,}
\centerline{\small \it and Department of Physics, Peking University, 
            Beijing 100871, People's Republic of China}
\begin{center}
\begin{minipage}{130mm}
\vskip0.6in
\begin{center}{\bf Abstract}\end{center}
{We present a calculation of the leading order QCD fragmentation functions 
for gluons to split into spin-triplet D-wave quarkonia. We apply them to
evaluate the gluon fragmentation contributions to inclusive$~^3D_J$ 
quarkonium production at large transverse momentum processes like the 
Tevatron and find that the D-wave quarkonia, especially the charmonium
$2^{--}$ state, could be observed through color-octet mechanism with present
luminosity. Since there are distinctively large gaps between the contributions of
two different ({\it i.e}, color-singlet and color-octet) quarkonium production
mechanisms, our results may stand as a unique test to NRQCD color-octet 
quarkonium production mechanism.}
\vskip 1cm
PACS number(s):$12.38.Bx$, $~13.87.Fh$, $~14.40.Gx$
\end{minipage}
\end{center}
\vfill\eject\pagestyle{plain}\setcounter{page}{1}

\begin{center}
\section{INTRODUCTION}
\end{center} 

The production and decays of quarkonium bound states have been under active 
studies in experiment ever since the first charmonium $1^{--}$ state, the 
$J/\psi$, was found twenty years before. The study of properties
of the bound states of heavy quarks has provided a wealth of information on 
both the properties of heavy quarks ($c$, $b$) themselves and quantum
chromodynamics since it stands on the very border between perturbative and
nonperturbative domains. Recently, the observations of the Collider Detector
at Fermilab (CDF) on the prompt charmonium production \cite{1p} has 
greatly stimulated progress in theoretical studies of quarkonium physics.

The conventional wisdom was that the dominant contributions to quarkonium 
production cross section at large transverse momentum ($P_T$) in $p\bar p$
collisions come from the QCD leading order diagrams, $i.e.$ the so called parton 
fusion processes. However, these calculations for prompt $J/\psi$ 
$(\psi^\prime)$ production did not reproduce all the aspects of the available
data \cite{2p}. It was pointed out by Braaten and Yuan \cite{3p} in 1993
that the dominant production mechanism at sufficiently large $P_T$ is the 
fragmentation of a parton produced with large transverse momentum, while 
formally this is of higher order in the strong coupling constant $\alpha_s$.
Unfortunately, even after including the fragmentation contributions, the 
predictions for the $\psi'$ production rate still falloff far below
the data \cite{4p}. This large discrepency between theory and experiment 
has called
in question the simple color-singlet model description for quarkonium \cite{5p} 
and suggests that a new paradigm for treating heavy quark-antiquark bound systems
that go beyond the color-singlet model might play an important role in the 
production of quarkonium at large $P_T$.

To this end, a factorization formalism has recently been performed by Bodwin,
Braaten, and Lepage \cite{6p} in the context of nonrelativistic quantum
chromodynamics (NRQCD), which provides a new framework to calculate the 
inclusive production and decay rates of quarkonia. In this approach, the 
calculations are organizeded in powers of $v$, the average velocity of the 
heavy quark (antiquark) in the meson rest frame, and in $\alpha_s$, the strong 
coupling constant. In NRQCD, quarkonium is not solely regarded as simply a 
quark-antiquark pair but rather a superposition of Fock states. The general
Fock state expansion starts as
\begin{eqnarray}
\label{a1}
|H(nJ^{PC})>&=&{\cal O}(1)|Q\bar{Q}(^{2S+1}L_J,\b 1)>\\ \nonumber
            &+&{\cal O}(v)|Q\bar{Q}(^{2S+1}(L\pm 1)_{J'},\b{8}) g>\\\nonumber
            &+&{\cal O}(v^2)|Q\bar{Q}(^{2S+1}L_J,\b{8}) g g > + \cdots\\ \nonumber
            &+&\cdots
\end{eqnarray}
where the angular momentum quantum numbers of the $Q\bar{Q}$ pairs within 
various Fock components are indicated in spectroscopic notation inside the 
brackets with color configuration of either $\b{1}$ or $\b{8}$.

The breakdown of color singlet model stems from its overlook of high Fock
components contributions to quarkonium production cross sections. The 
color-octet term in the gluon fragmentation to $J/\psi$($\psi'$) has been 
considered by Braaten and Fleming \cite{4p} to explain the $J/\psi$($\psi'$)
surplus problems discovered by CDF. Taking $<{\cal O}^{J/\psi}_8(^3S_1)>$ and
$<{\cal O}^{\psi'}_8(^3S_1)>$ as input parameters, the CDF surplus problems
for $J/\psi$ and $\psi'$ can be explained as the contributions of color-octet
terms due to gluon fragmentation.

Even though the color-octet mechanism has gained some successes in describing 
the production and decays of heavy quark bound systems 
\cite{6p}\cite{xx}\cite{7p},
it still needs more effort to go before finally setting its position and role in
heavy quarkonium physics. Therefore the most urgent task among others needs to 
do now is to confirm and identify the color-octet quarkonium signals.

While the first charmonium state, the $J/\psi$, has been found over twenty
years, $D$-wave states, given the limited experimental data, have received
less attention. However, this situation may be changed in both experimental 
and theoretical investigations. 
Experimentally, there are hopes of observing charmonium $D$-wave
states in addition to the $\psi(3770)$, in a high-statistic exclusive
charmonium production experiment\cite{8p}, and $b\bar{b}$~ $D$-wave states in
$\Upsilon$ radiative decays\cite{9p}. 

Recently, there is some clue for the $D$-wave
$2^{--}$ charmonium state in $E705$ $300$ GeV $\pi^\pm$- and proton- Li 
interaction experiment \cite{10p}. In this 
experiment there is an abnormal phenomenon
that in the $J/\psi\pi^+\pi^-$ mass spectrum, two peaks at 
$\psi(3686)$ 
and at $3.836~GeV$ (given to be the $2^{--}$ state) are observed and they
have almost the same
height. Obviously, 
this situation is difficult to explain based upon the color-singlet 
model. However, it might be explained 
with the NRQCD analysis. 
Of course, at energies in fixed
target experiments like $E705$, the color-octet gluon fragmentation dominance
may or may not be the case. Moreover, the strong signal of $J/\psi\pi^+\pi^-$
at $3.836 GeV$ observed by $E705$ is now questioned by other experiments
\cite{11p}. Nevertheless, if the $E705$ result is confirmed 
(even with a smaller
rate, say, by a factor of 3, for the signal at $3.836 GeV$), the color-octet
gluon fragmentation will perhaps provide a quite unique explanation for the
$D$-wave charmonium production. 
It does remind
us that in the NRQCD approach, as discussed in Ref. \cite{12p}, 
the production rates of $D$-wave heavy quarkonium
states may be as large as that of $S$-wave states as long as the color-octet
gluon production mechanism dominantes. 

Study shows that both {\bf LEP} and {\bf Tevatron}, especially the latter, 
are suitable grounds to find
the D-wave quarkonia \cite{12p} and to test the color-octet signals \cite{13p}.
In this paper we find that the divergences of the contributions between
color-singlet and
color-octet mechanisms in quarkonium $^3\!D_J$ production is enormous. The
rest of the paper is arranged as follows: In Sec. II,
we describe the formalism, give out the fragmentation functions of 
$g\rightarrow ^3D_J$ to leading order in $\alpha_s$
and further calculate the fragmentation probabilities of the gluon 
to D-wave quarkonium states. In Sec. III, we apply the fragmentation 
functions to evaluate the D-wave quarkonium production rates at {\bf Tevatron}
and close with some thoughts and discussions.

\begin{center} 
\section{FORMALISM}
\end{center} 

Fragmentation is the formation of a hadron within a jet produced by a parton 
(quark, antiquark or gluon) with large transverse momentum. It is a useful
concept because the probability for the formation of hadron within a jet is 
independent of the process that produces the parton that initiates the jet.
By now, the fragmentation functions of quark and gluon
splitting to S- and P-wave heavy quark bound states have been calculated
\cite{13p}\cite{14p}\cite{15p}\cite{16p}. The calculations of
fragmentation functions of quark to D-wave states \cite{17p} and gluon to
spin-singlet D-wave state $^1D_2$ have also been accomplished 
\cite{18p}. However, the study of gluon fragmentation
to color-singlet $^3D_J$ states, for its complexity, is still left behind.
Here we realize this goal.

In hard process, as Fig.1(a), the most important kinematic region for a virtual
gluon split with large $P_T$ is that the gluon is nearly on its massshell. 
Therefore, we can estimate the decay widths and the branching ratios by the  
following way \cite{19p}.

The decay widths of a virtual quark $Q^*$ to color-singlet quarkonium state
$^3D_J$ by gluon fragmentation can be evaluated via
\begin{eqnarray}
\Gamma(Q^*\rightarrow Q g^*; g^*\rightarrow ^3\!D_J~gg)=\int
\limits^s_{\mu^2_{min}}\!d\mu^2~\Gamma(Q^*\rightarrow Q g^*(\mu))\cdot 
P(g^*\rightarrow ^3\!D_J~gg),
\end{eqnarray}
where $s$ is the invariant mass squared of $Q^*$; $\mu$ is the virtuality 
of the gluon, and its minimum value squared $\mu_{min}^2=12 m_Q^2$ corresponding 
to the infrared cutoff as discussed below; 
$P$ is the decay distribution defined as
\begin{equation}
  P(g^*\rightarrow AX)\equiv\frac{1}{\pi\mu^3 }\Gamma(g^*\rightarrow AX).
\end{equation}

The general covariant procedure for calculating the production and decay
rates of heavy quark bound states may start from the Bethe-Salpeter(BS)
amplitudes in the nonrelativistic limit. At leading order in $\alpha_s$,
the amplitudes for $g^*\rightarrow ^3\!D_J~gg$ processes are
\begin{equation}
\label{4x} 
 {\cal A}=\int\frac{d^4 q}{(2\pi)^4}Tr\{{\cal O}(P,q) \chi(P,q)\}.
\end{equation}
Here $\chi(P,q)$ is the BS wave function of the bound states with
relative momentum $q$ between the heavy quarks,
while ${\cal O}(P,q)$ represents the rest of the matrix elements depicted in
Fig.1(a),
\begin{eqnarray}
\nonumber 
 {\cal O}(q)&=&-\frac{i}{4}\big\{\not\! \epsilon_2\frac{-\not\! k_1
 +\not\! k_2+\not\! k-2m_Q}{-(k_1-k)\cdot k_2}\not\! \epsilon
 \frac{-\not\! k_1+\not\! k_2-\not\! k-2m_Q}{-(k_2-k)\cdot k_1}
  \not\! \epsilon_1 \big\}\\
  &+& {\rm five~permutations~ of}~ k_1,~ k_2, ~-k~ {\rm and}~ \epsilon_1,~ 
  \epsilon_2,~ \epsilon~.
\end{eqnarray}
Here the $k_1,~k_2,~k$ and $\epsilon_1,~\epsilon_2,~\epsilon$ stand for the 
momenta and polarization vectors of the two outgoing final gluons and
the splitting gluon. Coupling constants and color matrices have been supressed
and contribute a factor
\begin{equation}
\sum\limits_{a,b,c}(\frac{1}{\sqrt{3}}g_s^3Tr\{T^aT^bT^c\})^2=
    \frac{5}{18}g_s^6
\end{equation}
to the production rates.

Under the instantaneous approximation with the negative energy projectors
being neglected, the BS wave function $\chi(P,q)$ may be expressed as
\begin{eqnarray}
\label{add6}
\chi(P,q)=\frac{i}{2 \pi}
\frac{P_0-E_1-E_2}{(p_{10} -E_1)(p_{20}-E_2)}\Phi(\vec{P},\vec{q}).
\end{eqnarray}
Here $P_0$ is the time component of the four momentum of the bound 
state; $ p_{10}$ and $p_{20}$ are the time 
components of the momenta of  quark and antiquark 
inside the meson, and $ E_1, E_2$ are their kinetic
energies. 
From the standard BS wave functions in the  
approximation that the negative energy projectors are omitted, 
the vector meson wave function can be projected out as :

\begin{equation}
 \Phi(\vec{P},\vec{q})=\frac{1}{M}\sum_{S_z m}
 \langle JM|1S_z Lm\rangle \Lambda^1_+ (\vec{p_1})\gamma_0 
  \not\!{e}(M+\not\!{P})\gamma_0\Lambda^2_- (\vec{p_2})\psi_{Lm}
         (\vec{P},\vec{q}),
\end{equation}
where $e$ is the polarization vector associated with the spin-triplet 
states. $\Lambda^1_+(\vec{p_1})$ and $\Lambda^2_-(\vec{p_2})$ 
are positive energy projection operators of quark and antiquark .
\begin{eqnarray}
  \label{add8}
  \Lambda^1_+(\vec{p_1})=\frac{E_1+\gamma_0 \vec{\gamma}\cdot\vec{p_1} + 
   m_1\gamma_0}{2E_1},~~~~
  \Lambda^2_-(\vec{p_2})=\frac{E_2-\gamma_0 \vec{\gamma}\cdot\vec{p_2} - 
  m_2\gamma_0}{2E_2}.
\end{eqnarray}
After taking the nonrelativistic approximation the bound state wave function 
may be further reduced. For D-wave quarkonium production and decay, the first 
nonzero term is proportional to the second order of the amplitude expansion
in powers of $q/M$:
\begin{eqnarray}
{\cal A}(P,q) = {\cal A}(P,0) + 
q_\alpha \frac{\partial {\cal A}(P,q)}{\partial q_\alpha}|_{q=0} +
\frac{1}{2} q_\alpha q_\beta\frac{\partial^2 {\cal A}(P,q)}{\partial q_\alpha
\partial q_\beta}|_{q=0} + \cdots. 
\end{eqnarray}
After integrating $q_\alpha q_\beta$ over $d^4 q $, the $^3D_J$ 
polarization tensor is related to its nonrelativistic wavefunction by
\begin{equation}
 \int\frac{d^3 q}{(2\pi)^3}q_{\alpha}q_{\beta}
 \psi_{2 m}(\vec{P},\vec{q})=e_{\alpha\beta}^{(m)}\sqrt{\frac{15}{8\pi}}
 R_2^{\prime\prime}(0),
\end{equation}
where the polarization tensor's label $m$ ranges over the helicity levels of 
the $L=2$ meson. For spin-singlet case $e_{\alpha\beta}^{(m)}$ is identified
with $e_{\alpha\beta}^{(J_z)}$, while for the spin-triplet case, 
using explicit Clebsch-Gorden coefficients, we have the following 
spin-orbit momentum coupling forms \cite{yy}, 
\begin{eqnarray}
\nonumber  
  \sum_{S_zm}\langle 1J_z|1S_z 2 m\rangle e_{\alpha\beta}^{(m)}
  e_\rho^{(s)}
 &=&-[{3\over 20}]^{1/2}[(g_{\alpha\rho}-\frac{p_\alpha p_\rho}{4m_Q^2})\epsilon_\beta^{(J_z)}
     +(g_{\beta\rho}-\frac{p_\beta p_\rho}{4m_Q^2})\epsilon_\alpha^{(J_z)}\\
\label{n1}  
  &~&-{2\over 3}(g_{\alpha\beta}-\frac{p_\alpha p_\beta}{4m_Q^2})\epsilon_\rho^{(J_z)}],\\
\label{n2}
\sum_{S_z m}\langle 2J_z|1S_z 2m\rangle e_{\alpha\beta}^{(m)}
  e_\rho^{(s)}
  &=&\frac{i}{2\sqrt{6}m_Q}(\epsilon_{\alpha\sigma}^{(J_z)}
  \epsilon_{\tau\beta\rho\sigma^{\prime}}p^\tau g^{\sigma\sigma^\prime}+
   \epsilon_{\beta\sigma}^{(J_z)}
   \epsilon_{\tau\alpha\rho\sigma^{\prime}}p^\tau g^{\sigma\sigma^\prime}),\\
\label{n3}  
  \sum_{S_z m}\langle 3J_z|1S_z 2m\rangle e_{\alpha\beta}^{(m)}
   e_\rho^{(s)}
    &=&\epsilon_{\alpha\beta\rho}^{(J_z)}.
\end{eqnarray}
Using Eqs. (\ref{n1})-(\ref{n3}) listed above, the amplitudes of Eq.(\ref{4x})
may be simplified and the averaged squared amplitudes may be obtained 
when suming up all polarizations of both the meson and gluons. Because the
results are lengthy, it is too tedious to write them all here. For the
convenience of reference, we just give the expression for the $^3D_1$ state
in the Appendix. Then, we have
\begin{eqnarray}
\label{tt}
\Gamma(g^*\rightarrow ^3\!D_J~gg)=\int dx_1 d x_2 \overline {\sum}|{\cal A}|^2,
\end{eqnarray}
where the kinematic variables are defined $x_1=\frac{2 k\cdot k_1}{\mu^2}$ and
$x_2=\frac{2 k\cdot k_2}{\mu^2}$. Furthermore, from the Eq.(\ref{tt})
we can get the expressions of decay distributions 
$P(g^*\rightarrow ^3\!D_J~gg)$. With them 
the fragmentation functions can be calculated straightforward
\begin{eqnarray}
\label{4y}
D_{g^*\rightarrow^3D_J}(z,2m_Q,s)=\frac{d\Gamma(Q^*\rightarrow ^3\!D_J~gg~Q)/dz}
{\Gamma(Q^*\rightarrow Q g )},
\end{eqnarray}
where $z\equiv\frac{2P\cdot k}{\mu^2}=2-x_1-x_2$. At high energy limit, the  
interaction energy $s$ goes up to infinity, then the definition of $z$ here 
is identical with that in Ref.\cite{16p} multiplied by a factor of two
and the fragmentation functions decouple from any specific gluon splitting
processes, which just reflects the universal spirit of fragmentation. 
The fragmetation function of 
Eq.(\ref{4y}) is evaluated at the renormalization scale $2m_Q$, which
corresponds to the minimum value of the invariant mass of the 
virtual gluon. 
In Fig.2 we display the variation curves of $D_{g^*\rightarrow^3D_J}(z,2m_c)$ 
versus $z$. After integrating over 
variable $z$, the fragmentation probabilities then read as
\begin{eqnarray}
\label{4}
P_{g^*\rightarrow^3D_J}=\frac{\Gamma(Q^*\rightarrow ^3\!D_J~gg~Q)}
{\Gamma(Q^*\rightarrow Q g )}.
\end{eqnarray}
Studies show \cite{16p} that the above method in extracting the gluon
fragmentaiton probabilites are equivalent to the method developed in 
Ref. \cite{3p}. 

The calculation of color-octet fragmentation functions in
$g^*\rightarrow{}^3\!D_J({}^3S_1,\b 8)$ processes, as shown in Fig.1(b), is
trivial. They may be obtained directly from color-octet
$g^*\rightarrow J/\psi({}^3\!S_1,\b 8)$ process \cite{4p},
\begin{eqnarray}
\label{5x}
D_{g^*\rightarrow^3\!D_J}(z,2m_Q)=\frac{\pi\alpha_s(2m_Q)}{24 m_Q^2}\delta(1-z)
<{\cal O}_8^{^3D_J}(^3S_1)>.
\end{eqnarray}
Therefore, the fragmentation probabilities are expressed as:
\begin{eqnarray}
\label{5}
P_{g^*\rightarrow^3\!D_J}=\frac{\pi\alpha_s(2m_Q)}{24 m_Q^2}
<{\cal O}_8^{^3D_J}(^3S_1)>.
\end{eqnarray}

\begin{center}
\section{RESULTS AND DISCUSSIONS}
\end{center} 

From  Eq.(\ref{4}) and (\ref{5}) we can estimate the quarkonium 
$^3D_J$ production rates at the {\bf Tevatron}. The color-singlet sector 
may be factorized into long distance and short distance terms. The former 
is, to leading order in $v^2$, proportional to the second derivative of 
the radial wave function at the origin, which may 
be determined from potential model calculations \cite{20p}. The latter  
can be calculated
from  perturbative QCD, and it involves the infrared divergence associated with
a soft gluon in the final states. 
In the numerical computation, we impose a lower 
cutoff $\Lambda$ on the energies of either gluons in the quarkonium rest frame.
As discussed in Ref. \cite{16p}, 
we choose $\Lambda=m_Q$ to avoid large logarithms
and the cutoff dependence of the color-singlet terms is cancelled by the 
$\Lambda$ dependence of the nonperturbative matrix elements 
$<{\cal O}_8^{^3D_J}(^3S_1)>$ of the corresponding color-octet terms.

The gluon fragmentation contributions to the production of quarkonium $^3D_J$ 
states at
large transverse momentum in any high energy process can be approximately 
obtained 
by multiplying the cross section for producing gluons with transverse momentum 
larger than $2m_c$ by appropriate fragmentaion probabilities \cite{7p}. Using
\cite{14p}\cite{20p}
$$m_c=1.5~GeV, m_b=4.9~GeV,~ \alpha_s(2m_c)=0.26,~ \alpha_s(2m_b)=0.19,$$
\vskip -0.9cm
\begin{eqnarray}
\label{aaaa}
|R^{\prime\prime}_{(c\bar c)}(0)|^2=0.015~ GeV^7,~
|R^{\prime\prime}_{(b\bar b)}(0)|^2=0.637~ GeV^7.
\end{eqnarray}
We obtain
\begin{eqnarray}
\label{7}
D^{(1)}_{g^*\rightarrow^3D_1(c\bar c)}=5.6\times 10^{-8},~
D^{(1)}_{g^*\rightarrow^3D_2(c\bar c)}=3.1\times 10^{-7},\nonumber\\
D^{(1)}_{g^*\rightarrow^3D_3(c\bar c)}=2.2\times 10^{-7},~
D^{(1)}_{g^*\rightarrow^3D_1(b\bar b)}=2.5\times 10^{-10},\nonumber\\
D^{(1)}_{g^*\rightarrow^3D_2(b\bar b)}=1.4\times 10^{-9},~
D^{(1)}_{g^*\rightarrow^3D_3(b\bar b)}=9.9\times 10^{-10}.
\end{eqnarray}

For gluon fragmentation color-octet processes, the fragmentation 
probabilities are proportional to the nonperturbative matrix elements 
$<{\cal O}_8^{^3D_J}(^3S_1)>$ which
have not been extracted out from experimental data, nor from the Lattice
QCD calculations. Based upon the NRQCD 
velocity scaling rules and the experimental clues discussed above, here we 
tentatively assume \cite{12p} 
\begin{eqnarray}
\label{9}
<{\cal O}_8^{^3D_2(c\bar c)}(^3S_1) > \approx 
<{\cal O}_8^{\psi'}(^3S_1)>=4.6\times 10^{-3}~GeV^3
\end{eqnarray}
(see Ref.\cite{xx}) 
and further extend this relation to the $b\bar b$ system \cite{xx}
\begin{eqnarray}
\label{10}
<{\cal O}_8^{^3D_2(b\bar b)}(^3S_1)> \approx 
<{\cal O}_8^{\Upsilon'}(^3S_1)>=4.1\times 10^{-3}~GeV^3.
\end{eqnarray}
The supposed relations (\ref{9}) and (\ref{10}) certainly possess uncertainties
to some extent, however from the calculated results below we are confident that 
it will not destroy the major conclusion of this paper.
From the approximate heavy quark spin symetry relation, we have
\begin{eqnarray}
\label{11}
<{\cal O}_8^{^3D_1}(^3S_1)> \approx \frac{3}{5}
<{\cal O}_8^{^3D_2}(^3S_1)> \approx \frac{5}{7} 
<{\cal O}_8^{^3D_3}(^3S_1)>
\end{eqnarray}
for both $b\bar{b}$ and $c\bar{c}$ systems.

Using Eqs.(\ref{5}), (\ref{aaaa}), and (\ref{9})-(\ref{11}), we readily have
\begin{eqnarray}
\label{12}
D^{(8)}_{g^*\rightarrow^3D_1(c\bar c)}=4.2\times 10^{-5},~
D^{(8)}_{g^*\rightarrow^3D_2(c\bar c)}=7.0\times 10^{-5},\nonumber\\
D^{(8)}_{g^*\rightarrow^3D_3(c\bar c)}=9.7\times 10^{-5},~
D^{(8)}_{g^*\rightarrow^3D_1(b\bar b)}=2.5\times 10^{-6},\nonumber\\
D^{(8)}_{g^*\rightarrow^3D_2(b\bar b)}=4.2\times 10^{-6},~
D^{(8)}_{g^*\rightarrow^3D_3(b\bar b)}=5.9\times 10^{-6}.
\end{eqnarray}
Comparing the above results (\ref{12}) with (\ref{7}), we come to an 
anticipated conclusion that  at the {\bf Tevatron} 
the gluon fragmentation probabilities through color-octet intermediates
to spin-triplet D-wave charmonium and bottomnium states are over $2\sim 4$ 
orders of 
magnitude larger than that of color-singlet processes. 
As a result, the production rates of $^3\!D_J$ states are about the same
amount as $\psi'$ and $\Upsilon(2S)$ production rates. Compared with the 
$\psi'$ production at the {\bf Tevatron}, the gluon fragmentation color-octet
process plays an even more important role in the $^3D_J$ quarkonium production,
and it also 
gives production probabilities larger than the quark fragmentation 
process \cite{17p}.

Among the three triplet states of $D$-wave charmonium, $^3D_2$ is the 
most promising candidate
to discover firstly. Its mass falls in the range of $3.810\sim 3.840~GeV$ in the 
potential model calculation\cite{l1}, that is above the
$D{\bar D}$ threshold but below the $D{\bar D}^*$
threshold. However the parity conservation forbids it decaying into $D\bar D$. 
It, therefore, is a narrow resonance. Its main decay modes are expected to be,
\begin{equation}
^3\!D_2\rightarrow J/\psi\pi\pi,~~~^3\!D_2\rightarrow ^3\!P_J \gamma(J=1,2),~~~
^3D_2\rightarrow 3g.
\end{equation}
We can estimate the hadronic transition 
rate of $^3D_2\rightarrow J/\psi\pi^+\pi^-$
from the Mark III data for 
$\psi(3770)\rightarrow J/\psi\pi^+\pi^-$\cite{zhu} and
the QCD multipole expansion theory\cite{yan}\cite{ky}. The Mark III data
give\cite{zhu}
$\Gamma(\psi(3770)\rightarrow J/\psi\pi^+\pi^-)=(37\pm 17\pm 8)~ keV~~~
{\rm or}~~~(55\pm 23\pm 11)~ keV$
(see also Ref.\cite{ky}). Because the $S-D$ mixing angle for $\psi(3770)$ and
$\psi(3686)$ is expected to be small 
(say, $-10^\circ$, see Ref.\cite{ding} for
the reasoning), 
the observed $\psi(3770)\rightarrow J/\psi\pi^+\pi^-$ transition
should dominantly come from the $^3D_1\rightarrow J/\psi\pi^+\pi^-$ transition,
which is also compatible with the multipole expansion estimate\cite{ky}.
Then using the relation\cite{yan}
$$
d\Gamma(^3\!D_2\rightarrow ^3\!S_1 2\pi)=d\Gamma(^3\!D_1\rightarrow ^3\!S_1 2\pi)
$$
and taking the average value of 
the $\Gamma(\psi(3770)\rightarrow J/\psi\pi^+\pi^-)$
from the Mark III data, we may have
\begin{equation}
\label{y1}
\Gamma(^3D_2\rightarrow J/\psi \pi^+\pi^-)=\Gamma(^3D_1\rightarrow J/\psi \pi^+\pi^-)
\approx 46~ keV.
\end{equation}
For the E1 
transition $3\!D_2\rightarrow ^3\!P_J\gamma(J=1,2)$, using the potential
model with relativistic effects being considered\cite{dchao}, we find
\begin{equation}
\label{y2}
\Gamma(^3D_2 \rightarrow \chi_{c1}\gamma)=250~ keV,~~~
\Gamma(^3D_2 \rightarrow \chi_{c2}\gamma)=60~ keV,
\end{equation}
where the mass of $^3D_2$ is set to be $3.84 GeV$. 
As for the $^3D_2\rightarrow
3g$ annihilation decay, an estimate gives\cite{be}
\begin{equation}
\label{y3}
\Gamma(^3D_2\rightarrow 3g)=12~ keV
\end{equation}
From (\ref{y1}), (\ref{y2}), and (\ref{y3}), we find
\begin{eqnarray}
\nonumber
\Gamma_{tot}(^3D_2)&\approx & \Gamma(^3D_2\rightarrow J/\psi \pi\pi)
+\Gamma(^3D_2 \rightarrow \chi_{c1}\gamma)+\Gamma(^3D_2 \rightarrow \chi_{c2}\gamma)
+\Gamma(^3D_2\rightarrow 3g)\\
&\approx &390~ keV,
\end{eqnarray}
and
\begin{equation}
\label{17}
B(^3D_2\rightarrow J/\psi \pi^+ \pi^{-} )\approx 0.12.
\end{equation}
Considering all the uncertainties this estimate is expected to hold within
50\%.
Compared (\ref{17}) with $B(\psi^\prime\rightarrow J/\psi \pi^+\pi^-)=0.324\pm 0.026$,
the branching ratio of $^3D_2\rightarrow J/\psi \pi^+\pi^-$ is only smaller
by a factor of 3, and therefore the decay mode of $^3D_2\rightarrow J/\psi \pi^+\pi^-$
could be observable at {\bf Tevatron}.

The $^3D_1$ $c\bar c$ state $\psi(3770)$ could also detected via 
$^3D_1\rightarrow D \bar D$.
The other states, including the $^3D_3(c\bar c)$ and 
$^3D_J(b\bar b)$ are perhaps difficult to detect for reasons of either
more decay modes or smaller production rates.

In conclusion, we have calculated the fragmentation functions and 
fragmentation probabilities
of the gluon to $^3D_J$ charmonium and bottomonium states in both color-singlet 
and color-octet processes 
with certain numerical assumptions (e.g. Eq.(\ref{9})). 
The results can also be used in other hard gluon 
fragmentation processes because of the universality of the fragmentation
functions. The study shows that, because charmonium $^3D_2$ state may have a  
production rate as large as that of $\psi^\prime$ at the {\bf Tevatron} 
through color-octet production mechanism, the charmonium $^3D_2$ state as a 
most promising candidate to discover should be observable at the 
{\bf Tevatron} with 
present luminosity, even the assumption of Eq.(\ref{9}) with an error of 10 
times off the exact case. On the other hand, since 
the calculated results show
that the color-singlet and the color-octet contributions diverge 
enormously, this will also present a crucial 
test for the color-octet mechanism.
the $^3D_J$ bottomonium states may have less strong signals to be detected
comparing with the $^3D_J$ charmonium states because of their 
small production rates.

\vskip 1cm
\begin{center}
\bf\large\bf{ACKNOWLEDGEMENTS}
\end{center}

This work was supported in part by the National Natural Science Foundation
of China, the State Education Commission of China and the State Commission
of Science and Technology of China.

\newpage
\appendix
\centerline{\bf APPENDIX}
The expression of the averaged squared amplitude of the process $g^*\rightarrow
{}^3D_1 gg$ 
\begin{equation}
\nonumber
\overline{\sum}|{\cal A}|^2=\frac{25\alpha_s^2\mu|R_D^{\prime\prime}(0)|^2}{2^{15}\pi m_c^7}
        \frac{128 d \sum\limits_{i=0}^{12} f_i d^i}
        {15(1-d-x_1)^5(1-d-x_2)^5(x_1+x_2)^6},
\end{equation}
where $d=M^2/\mu^2$ and the functions $f_i$ are defined as
\begin{eqnarray}
\nonumber
f_0&=&(x_1 - 1)^4 (x_2 - 1)^4[x_1^5 (x_2 + 17) + x_1^4 (4 x_2^2 + 53 x_2 - 177)
   + x_1^3 (6 x_2^3 + 74 x_2^2 - 420 x_2 \\
\nonumber
   &~&+ 576) + x_1^2 (4 x_2^4 + 74 x_2^3 - 486 x_2^2 + 1216 x_2 - 928) 
     +x_1 (x_2^5 + 53 x_2^4 - 420 x_2^3 + 1216 x_2^2\\
\nonumber
   &~& - 1600 x_2 +768) 
   +17 x_2^5 - 177 x_2^4 + 576 x_2^3 - 928 x_2^2 + 768 x_2 - 256],\\
\nonumber
f_1&=&(x_1 - 1)^3 (x_2 - 1)^3 [10 x_1^8 + 4x_1^7 (15 x_2 + 1) + x_1^6 (170 x_2^2 + 39 x_2 - 
        141) + x_1^5 (300 x_2^3 \\
\nonumber
    &~& - 133 x_2^2 - 616 x_2 - 271)+ x_1^4 (360 x_2^4 - 566 x_2^3 - 803 x_2^2 - 587
        x_2 + 3162) + x_1^3 (300 x_2^5\\ 
\nonumber
    &~&- 566 x_2^4 - 656x_2^3 + 106 x_2^2 + 8336 x_2 - 8696) + x_1^2 (170 x_2
        ^6 - 133 x_2^5 - 803 x_2^4 + 106 x_2^3 \\
\nonumber
    &~&+ 9772x_2^2 - 21480 x_2 + 12720) + x_1 (60 x_2^7 + 39 x_2^6
        - 616 x_2^5 - 587 x_2^4 + 8336 x_2^3 \\
\nonumber
    &~&- 21480 x_2^2 + 24160 x_2 - 9856) + 10 x_2^8 + 4 x_2^7 - 141 x_2^6 - 271 x_2^5 +
        3162 x_2^4 - 8696 x_2^3 \\
\nonumber
    &~&+ 12720 x_2^2 - 9856 x_2 + 3136],\\
\nonumber
f_2&=&(x_1 - 1)^2 (x_2 - 1)^2[- 50 x_1^9 + 2 x_1^8 ( - 125 x_2 + 14) + 2 x_1^7 ( - 320 x_2^2 + 479 
        x_2 - 97) + x_1^6 ( 1951\\
\nonumber
    &~& + 3593 x_2^2 - 5120 x_2 - 1120 x_2^3 1951) + 2
         x_1^5 ( - 730 x_2^4 + 2819 x_2^3 - 8038 x_2^2+ 7736 x_2 - 160)\\ 
\nonumber
    &~& + x_1^4 ( - 1460 x_2^5 + 5950 x_2^4 - 24146 x_2^3 + 44649 x_2^2 - 11984 
        x_2 - 17579) + 2 x_1^3 ( - 560 x_2^6  \\
\nonumber
    &~&+ 2819 x_2^5- 12073 x_2^4 +
        31416 x_2^3 - 22632 x_2^2 - 24258 x_2 + 26192) + x_1^2 ( 3593 x_2^6- 640 x_2^7 \\ 
\nonumber
    &~&- 16076 x_2^5 + 44649 x_2^4 - 45264 x_2^3 - 57330
        x_2^2 + 146496 x_2 - 75496) + 2 x_1 ( 479 x_2^7- 125 x_2^8 \\
\nonumber
    &~&- 2560 x_2^6 + 7736 x_2^5 - 5992 x_2^4 - 24258 x_2^3 + 73248 x_2^2 -
        76552 x_2 + 27952) - 50 x_2^9 + 28 x_2^8\\
\nonumber
    &~& - 194 x_2^7 + 1951 x_2^6 - 320 x_2^5 - 17579 x_2^4 + 52384 x_2^3 
        - 75496 x_2^2 + 55904 x_2 - 16832],\\
\nonumber
f_3&=&2 (x_1 - 1) (x_2 - 1)[67 x_1^{10} + x_1^9 (277 x_2 - 251) + 3 x_1^8 
        (328 x_2^2 - 618 x_2 + 405)+ x_1^7 (2780 x_2^3 \\
\nonumber
&~&- 5699 x_2^2 + 3156 x_2 - 513) + x_1^6 (5153 x_2^4 - 12461 x_2^3 
        - 1731 x_2^2 + 18247 x_2 - 8577)\\ 
\nonumber
&~&+ x_1^5 (6294 x_2^5 - 19575 x_2^4 - 11940 x_2^3 + 89313 x_2^2 - 82353 x_2 + 15559) + 
        x_1^4 (5153 x_2^6\\ 
\nonumber
&~&- 19575 x_2^5 - 16760 x_2^4 + 178137 x_2^3 - 272631x_2^2 + 116247 x_2 
        + 11894) + x_1^3 (2780 x_2^7  \\
\nonumber
&~&- 12461 x_2^6- 11940 x_2^5 + 178137 x_2^4 - 401134 x_2^3 + 303362 x_2^2 
        + 20372x_2 - 79388) \\
\nonumber
&~&+ x_1^2 (984 x_2^8 - 5699 x_2^7 - 1731 x_2^6 + 89313 x_2^5 - 272631 x_2^4 
        + 303362 x_2^3 + 10876 x_2^2 \\
\nonumber
&~&- 249684 x_2 + 125056) + x_1 (277 x_2^9- 1854 x_2^8 + 3156 x_2^7 
        + 18247 x_2^6 - 82353 x_2^5\\
\nonumber
&~& + 116247 x_2^4 + 20372 x_2^3 - 249684 x_2^2 + 266656 x_2- 91040) 
        + 67 x_2^{10} - 251 x_2^9 + 1215 x_2^8   \\
\nonumber
&~&- 513 x_2^7- 8577 x_2^6+ 15559 x_2^5 + 11894 x_2^4 - 79388 x_2^3 
        + 125056 x_2^2 - 91040 x_2 + 26080,\\
\nonumber
f_4&=&98 x_1^{11} + 2 x_1^{10} (629 x_2 - 640) + x_1^9 (6804 x_2^2 - 13203 x_2 
        + 6503) + x_1^8 (23612 x_2^3 \\
\nonumber
&~&- 63506 x_2^2 + 56515 x_2 - 17151) + 2 x_1^7 (26637 x_2^4 - 90183 x_2^3 
        + 97672 x_2^2 - 32167 x_2 \\
\nonumber
&~&- 1868) + 2x_1^6 (39621 x_2^5 - 167139 x_2^4 + 205992 x_2^3 - 2891 x_2^2 
        - 135604 x_2 + 59500) \\
\nonumber
&~&+ 2 x_1^5 (39621 x_2^6 - 205399 x_2^5 + 296235 x_2^4 + 142371 x_2^3 
        - 702184 x_2^2 + 558303 x_2 - 127486) \\
\nonumber
&~&+2 x_1^4 (26637 x_2^7 - 167139 x_2^6 + 296235 x_2^5 + 246477 x_2^4 
        - 1484408 x_2^3 + 1835580 x_2^2 \\ 
\nonumber
&~&- 839002 x_2+ 84843) + 2 x_1^3 (11806 x_2^8 - 90183 x_2^7 
        + 205992 x_2^6 + 142371 x_2^5 - 1484408 x_2^4\\ 
\nonumber
&~&+ 2689458 x_2^3 - 1978824 x_2^2 + 403532 x_2 + 100128) 
        + 2 x_1^2 (3402 x_2^9 - 31753 x_2^8 + 97672 x_2^7 \\
\nonumber
&~&- 2891 x_2^6 - 702184 x_2^5 + 1835580 x_2^4 - 1978824 x_2^3 
        + 640546 x_2^2 + 380960 x_2 -242436)  \\
\nonumber
&~&+ x_1 (1258 x_2^{10}- 13203 x_2^9 + 56515 x_2^8 - 64334 x_2^7 
        - 271208 x_2^6 + 1116606 x_2^5 \\
\nonumber
&~&- 1678004 x_2^4 + 807064 x_2^3 + 761920 x_2^2 - 1085712 x_2 + 369120) 
        + 98 x_2^{11} - 1280 x_2^{10} \\
\nonumber
&~&+ 6503 x_2^9 - 17151 x_2^8 - 3736 x_2^7 + 119000 x_2^6 - 254972 x_2^5
        + 169686 x_2^4 + 200256 x_2^3 \\
\nonumber
&~&- 484872 x_2^2 + 369120 x_2 - 102720,\\
\nonumber
f_5&=&490 x_1^{10} + x_1^9 (4870 x_2 - 4351) + x_1^8 (25096 x_2^2 - 42443 x_2 
        +17526) + 2 x_1^7 (38960 x_2^3\\ 
\nonumber
&~&- 93761 x_2^2 + 63537 x_2 - 8870) + 2 x_1^6 (75695 x_2^4 - 240989 x_2^3 
        + 207627 x_2^2 + 12406 x_2\\ 
\nonumber
&~&- 54294) + 2 x_1^5 (94186 x_2^5 - 383133 x_2^4 + 407123 x_2^3 
        + 193818 x_2^2 - 526995 x_2 + 209916)\\ 
\nonumber
&~&+ 2 x_1^4 (75695 x_2^6 - 383133 x_2^5 + 507036 x_2^4 + 480502 x_2^3 
        - 1722286 x_2^2 + 1369912 x_2 \\
\nonumber
&~&- 320530) + 2 x_1^3 (38960 x_2^7 - 240989 x_2^6 + 407123 x_2^5 
        + 480502 x_2^4 - 2515042 x_2^3 \\
\nonumber
&~&+ 3194364 x_2^2 - 1563232 x_2 + 196056) + 2 x_1^2 (12548 x_2^8 
        - 93761 x_2^7 + 207627 x_2^6 \\
\nonumber
&~&+ 193818 x_2^5 - 1722286 x_2^4 + 3194364 x_2^3 - 2511804 x_2^2 
        +648200 x_2 + 70608) + x_1 (4870 x_2^9 \\
\nonumber
&~&- 42443 x_2^8 + 127074 x_2^7 + 24812 x_2^6 - 1053990 x_2^5 
        + 2739824 x_2^4 - 3126464 x_2^3 \\
\nonumber
&~&+ 1296400 x_2^2 + 360384 x_2 - 330624) + 490 x_2^{10} -4351 x_2^9 
        + 17526 x_2^8 - 17740 x_2^7 \\
\nonumber
&~&- 108588 x_2^6 + 419832 x_2^5- 641060 x_2^4 + 392112 x_2^3 
        + 141216 x_2^2 - 330624 x_2 + 131712,\\
\nonumber
f_6&=&940 x_1^9 + x_1^8 (9380 x_2 - 6619) + 2 x_1^7 (23754 x_2^2 
        - 33091 x_2 + 8148) + 2 x_1^6 (67062 x_2^3 \\
\nonumber
&~&- 142303 x_2^2 + 55480 x_2 + 22478) + 2 x_1^5 (111064 x_2^4 
        - 327485 x_2^3 + 178620 x_2^2 + 209904 x_2 \\
\nonumber
&~&- 168145) + 2 x_1^4 (111064 x_2^5 - 429639 x_2^4 + 310024 x_2^3 
        + 667150 x_2^2 - 1059801 x_2 + 386365)\\ 
\nonumber
&~&+ 2 x_1^3 (67062 x_2^6 -327485 x_2^5 + 310024 x_2^4 + 968408 x_2^3 
        - 2432902 x_2^2 + 1859180 x_2 \\
\nonumber
&~&- 431760) + 2 x_1^2 (23754 x_2^7 - 142303 x_2^6 + 178620 x_2^5 
        + 667150 x_2^4 - 2432902 x_2^3 \\
\nonumber
&~&+ 2981054 x_2^2 - 1469776 x_2 + 192024) + 2 x_1 (4690 x_2^8 
        - 33091 x_2^7 + 55480 x_2^6 + 209904 x_2^5\\ 
\nonumber
&~&- 1059801 x_2^4 + 1859180 x_2^3 - 1469776 x_2^2 +388080 x_2 
        + 45024) + 940 x_2^9 - 6619 x_2^8 \\
\nonumber
&~&+ 16296 x_2^7 + 44956x_2^6 - 336290 x_2^5 + 772730 x_2^4 
        - 863520 x_2^3 + 384048 x_2^2 \\
\nonumber
&~&+ 90048 x_2 - 104832,\\
\nonumber
f_7&=&2[410 x_1^8 + 3 x_1^7 (1849 x_2 - 804) + 2 x_1^6 (14361 x_2^2 
        - 15578 x_2- 1348) + x_1^5 (71493 x_2^3, \\
\nonumber
&~&- 131020 x_2^2 - 7327 x_2 + 72719) + x_1^4 (95816 x_2^4 - 253812 x_2^3 
        - 3044 x_2^2 + 426335 x_2  \\
\nonumber
&~&- 254736)+ x_1^3 (71493 x_2^5 - 253812 x_2^4 + 582 x_2^3 + 950322 x_2^2 
        - 1194424 x_2 + 406920) \\
\nonumber
&~&+ 2 x_1^2 (14361 x_2^6 - 65510 x_2^5 - 1522 x_2^4 + 475161 x_2^3 
        - 951640 x_2^2 + 691660 x_2 - 157488)\\ 
\nonumber
&~&+ x_1 (5547 x_2^7 - 31156 x_2^6 - 7327 x_2^5 + 426335 x_2^4 
        - 1194424 x_2^3 + 1383320 x_2^2 \\
\nonumber
&~&- 661824 x_2 + 77952) + 410 x_2^8 - 2412 x_2^7- 2696 x_2^6 
        + 72719 x_2^5 - 254736 x_2^4 \\
\nonumber
&~&+ 406920 x_2^3 - 314976 x_2^2 + 77952 x_2 + 19392],\\
\nonumber
f_8&=&454 x_1^7 + 2 x_1^6 (5231 x_2 - 879) + x_1^5 (49146 x_2^2 
        - 37439 x_2 -29865) + x_1^4 (98210 x_2^3\\ 
\nonumber
&~&- 138246 x_2^2 - 144901 x_2 + 193291) + 2 x_1^3 (49105 x_2^4 
        - 102277 x_2^3 - 148785 x_2^2 + 435590 x_2 \\
\nonumber
&~&- 221376) + 2 x_1^2 (24573 x_2^5 - 69123 x_2^4 - 148785 x_2^3 
        + 686625 x_2^2 - 744064 x_2 + 240264) \\
\nonumber
&~&+ x_1 (10462 x_2^6 - 37439 x_2^5 - 144901 x_2^4 + 871180 x_2^3 
        - 1488128 x_2^2 + 1017888 x_2 -221760)\\ 
\nonumber
&~&+ 454 x_2^7 - 1758 x_2^6 - 29865 x_2^5 + 193291 x_2^4 
        - 442752 x_2^3 + 480528 x_2^2 - 221760 x_2 + 13440,\\
\nonumber
f_9&=&550 x_1^6 + x_1^5 (8230 x_2 + 637) + x_1^4 (28258 x_2^2 - 6151 x_2 
        - 38526) + 2 x_1^3 (20578 x_2^3 \\
\nonumber
&~&- 10371 x_2^2 - 81952 x_2 + 71372) + 2 x_1^2 (14129 x_2^4 - 10371 x_2^3 
        - 126946 x_2^2 + 234980 x_2 \\
\nonumber
&~&- 106504) + x_1 (8230 x_2^5 - 6151 x_2^4 - 163904 x_2^3 
        + 469960 x_2^2 - 451168 x_2 + 140480)\\ 
\nonumber
&~&+ 550 x_2^6 + 637 x_2^5 - 38526x_2^4 + 142744 x_2^3 
        - 213008 x_2^2 + 140480 x_2 - 25280,\\
\nonumber
f_{10}&=&572 x_1^5 + x_1^4 (4036 x_2 + 2537) + 4 x_1^3 (2288 x_2^2 
        + 2455 x_2 - 6348) + 2 x_1^2 (4576 x_2^3 \\
\nonumber
&~&+ 7379 x_2^2 - 40504 x_2 + 27420) + 4 x_1 (1009 x_2^4 + 2455 x_2^3 
        - 20252 x_2^2 + 28844 x_2 - 12392) \\
\nonumber
&~&+ 572 x_2^5 + 2537 x_2^4 - 25392 x_2^3 + 54840 x_2^2 - 49568 x_2 + 14144,\\
\nonumber
f_{11}&=&4 [51 x_1^4 + 2 x_1^3 (101 x_2 + 245) + 2 x_1^2 (151 x_2^2 
        + 751 x_2 - 928) + 2 x_1 (101 x_2^3 \\
\nonumber
&~&+ 751 x_2^2 - 1920 x_2 + 1152) + 51 x_2^4 + 490x_2^3 - 1856 x_2^2 
        + 2304 x_2 - 976],\\
\nonumber
f_{12}&=&8 [51 x_1^2 + 6 x_1 (17 x_2 - 14) + 51 x_2^2 - 84 x_2 + 56].
\end{eqnarray}

\newpage
\centerline{\bf \large Figure Captions}
\vskip 2cm
\noindent
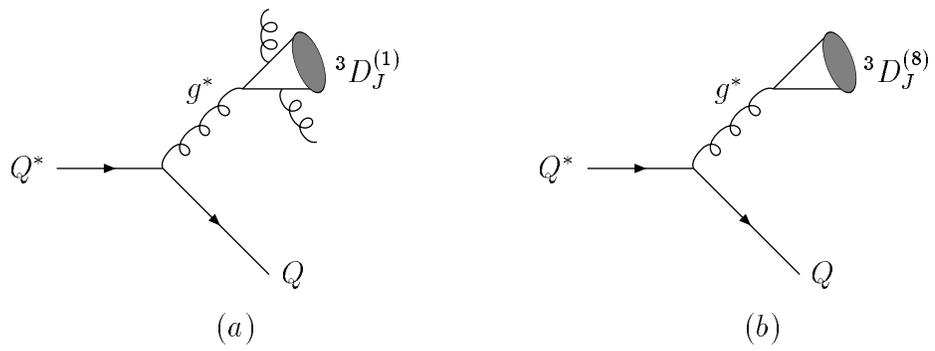
Fig.1. Virtual gluon fragmentation processes (a) gluon fragments to 
$^3\!D_J$ via color-singlet process, (b) gluon fragments to 
$^3\!D_J$ via color-octet process.
\vskip 0.2cm
\noindent
Fig.2  The variation of charmonium fragmetation functions 
$D_{(g\rightarrow ^3\! D_J)}(z, 2 m_c)$ versus $z$.

\newpage
\begin{references}
\bibitem{1p} CDF collaboration, F. Abe {\it et al}., Phys. Rev. Lett. {\bf 69}, 
   3704 (1992); Phys. Rev. Lett. {\bf 71}, 2537 (1993). 
\bibitem{2p} E. Braaten {\it et al.}, Phys. Lett. B{\bf 333}, 548 (1994);
        M. Cacciari and M. Greco, Phys. Rev. Lett. {\bf 73} (1994) 1586;
        D. P. Roy and K. Sridhar, Phys. Lett. B{\bf 339} (1994) 141.
\bibitem{3p} E. Braaten and T.C. Yuan, Phys. Rev. Lett. {\bf 71}, 1673(1993).
\bibitem{4p} E. Braaten and S. Fleming, Phys. Rev. Lett. {\bf 74}, 3327 (1995).
\bibitem{5p} For a recent review, see: G. A. Schuler, 
   Report No. CERN-TH 7170/94 (unpublished).
\bibitem{6p} G. T. Bodwin, E. Braaten, and G. P. Lepage, Phys. Rev.
 D{\bf 51}, 1125 (1995).
\bibitem{xx} P. Cho and K. Leibovich, Phys. Rev. D{\bf 53}, 150 (1996);
   {\it ibid}, D{\bf 53}, 6203 (1996).
\bibitem{7p} E. Braaten, S. Fleming, and T. C. Yuan, 
   Report No. hep-ph/9602374 (unpublished). 
\bibitem{8p} See, e.g. E. Menichetti, in Proceedings of the First Workshop 
   on Antimatter Physics at Low Energy, Batavia, Illinois, 1986, edited 
   by B. E. Bonner and L. S. Pinsky (Frmilab, Batavia, Illinois, 1986), P.95.
\bibitem{9p} J. L. Rosner, in {\it Particles and Fields 3}, 
   proceedings of the Banff Summer Institute (CAP) 1988, Banff, Alberta, 1988, edited by A. N. Kamal
   and F. C. Khanna (World Scientific, Singapore, 1989), P.395.
\bibitem{10p} L. Antoniazzi {\it et al.}, Phys. Rev. D{\bf 50}, 4258 (1994).
\bibitem{11p} $E672$ and $E706$ Collaboration,
        A. Gribushin {\it et al.}, Phys. Rev. D{\bf 53}, 4723 (1996).
\bibitem{12p} C. F. Qiao, F. Yuan, and K. T. Chao, Report No. hep-ph/9609284 
   (to appear on Phys. Rev. D).
\bibitem{13p} K. Cheung, W.-Y. keung, and T. C. Yuan, Phys. Rev. Lett {\bf 76}, 877 (1996).
\bibitem{14p} E. Braaten, K. Cheung, and T. C. Yuan, Phys. Rev. D{\bf 48}, 4230(1993).
\bibitem{15p} Y.-Q. Chen, Phys. Rev. D{\bf 48}, 5181 (1993).
\bibitem{16p} E. Braaten and T. C. Yuan, Phys. Rev. D{\bf 50}, 3176 (1994).
\bibitem{17p} K. Cheung and T.C. Yuan, Phys. Rev. D{\bf 53}, 3591 (1996).
\bibitem{18p} P. Cho and M. B. Wise, Phys. Rev. D{\bf 51 }, 3352 (1995).
\bibitem{19p} K. Hagiwara, A. D. Martin, and W. J. Stirling, Phys. Lett. B267, 527 (1991).
\bibitem{yy} L. Bergstr\"om, H. Grotch, and R. W. Robinett, Phys. Rev. 
 D{\bf 43}, 2157 (1991).
\bibitem{20p} E. J. Eichten and C. Quigg, Phys. Rev. D{\bf 52}, 1726 (1995).
\bibitem{l1}  W. Kwong, J. Rosner, and C. Quigg, Annu. Rev. Nucl. Part. Phys.
 {\bf 37}, 343 (1987); S. Godfrey and 
 N. Isgur, Phys. Rev. D{\bf 32}, 189 (1985).
\bibitem{zhu} Y. N. Zhu, Ph. D. thesis, Report No. CALT-68-1513, 1988.
\bibitem{yan} T. M. Yan, Phys. Rev. D{\bf 22}, 1652 (1980).
\bibitem{ky} Y. P. Kuang and T. M. Yan, Phys. Rev. D{\bf 41}, 155 (1990).
\bibitem{ding} Y. B. Ding, D. H. Qin, and K. T. Chao, Phys. Rev. D{\bf 44},
        3562 (1991).
\bibitem{dchao} K. T. Chao, Y. B. Ding, and D. H. Qin, Phys. Lett. B{\bf 381},
        282 (1993).
\bibitem{be} L. Bergstr\"om and P. Ernstr\"om, Phys. Lett. B{\bf 267}, 111 (1991).
\end{references}
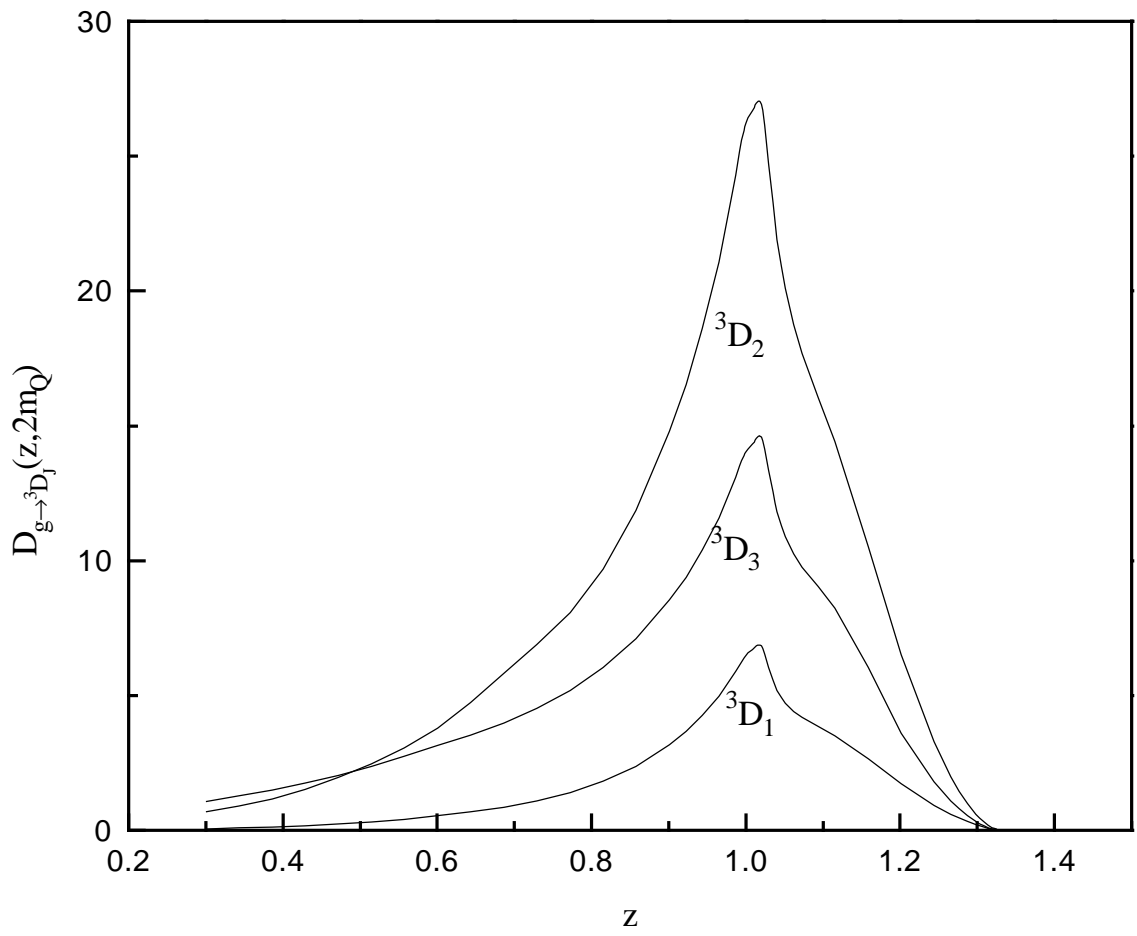
\end{document}